\documentclass{aa}  

\usepackage{graphicx,natbib}
\usepackage{txfonts}
\begin{document}

   \title{DGSAT: Dwarf Galaxy Survey with Amateur Telescopes}

   \subtitle{I. Discovery of low surface brightness systems around nearby spiral galaxies}

   \author{B. Javanmardi
          \inst{1,2,3}\fnmsep\thanks{Member of the International Max Planck Research School (IMPRS) for Astronomy and Astrophysics at the Universities of Bonn and Cologne.}\fnmsep\thanks{\email{behnam@astro.uni-bonn.de}},
          D. Martinez-Delgado
          \inst{4},
          P. Kroupa
          \inst{3}
          \and
          C. Henkel\inst{2,5}
          \and
          K. Crawford \inst{6} 
          \and   
K. Teuwen \inst{7}
\and      
R. J. Gabany \inst{8}
\and   
M. Hanson  \inst{9}
\and
          T. S. Chonis \inst{10}
\and    
F. Neyer    \inst{11}    
          }

   \institute{Argelander Institut für Astronomie der Universität Bonn, Auf dem Hügel 71, Bonn, D-53121, Germany
   \and
   Max-Planck-Institut für Radioastronomie, Auf dem Hügel 69, Bonn, D-53121, Germany
   \and
             Helmholtz-Institut für Strahlen und Kernphysik, Nussallee 14-16, Bonn, D-53115, Germany
         \and
             Astronomisches Rechen-Institut, Zentrum für Astronomie der Universitat Heidelberg, Monchhofstr. 12-14, 69120 Heidelberg, Germany
         \and
              Astronomy Department, Faculty of Science, King Abdulaziz University, PO Box 80203, 21589, Jeddah, Saudi Arabia
         \and
         Rancho del Sol Observatory, Modesto, California, USA
         \and
         Remote Observatories Southern Alpes, Verclause, France 
         \and
         Black Bird Observatory, 5660 Brionne Drive, San Jose, CA 95118, USA 
         \and
          Doc Greiner Research Observatory-Rancho Hidalgo, Animas, New Mexico, USA
          \and
         Department of Astronomy, University of Texas at Austin, 2515 Speedway, Stop C1400, Austin, TX 78712, USA
          \and
          ETH Zurich, Institute of Geodesy and Photogrammetry, 8093  Zurich, Switzerland
             }

   \date{Submitted: 13 November 2015 / Accepted: 01 February 2016}

 
  \abstract
   {We introduce the Dwarf Galaxy Survey with Amateur Telescopes (DGSAT) project and report the discovery of eleven low surface brightness (LSB) galaxies in the fields of the nearby galaxies NGC 2683, NGC 3628, NGC 4594 (M104), NGC 4631, NGC 5457 (M101), and NGC 7814.}
   {The DGSAT project aims to use the potential of small-sized telescopes to probe LSB features around large galaxies and to increase the sample size of the dwarf satellite galaxies in the Local Volume.}
   {Using long exposure images, fields of the target spiral galaxies are explored for extended low surface brightness objects. After identifying dwarf galaxy candidates, their observed properties are extracted by fitting models to their light profiles.}
   { We find three, one, three, one, one, and two new LSB galaxies in the fields of NGC 2683, 3628, 4594, 4631, 5457, and 7814, respectively. In addition to the newly found galaxies, we analyse the structural properties of nine already known galaxies. All of these 20 dwarf galaxy candidates have effective surface brightnesses in the range $25.3\lesssim\mu_{e}\lesssim28.8$ mag.arcsec$^{-2}$ and are fit with Sersic profiles with indices $n\lesssim 1$. Assuming that they are in the vicinity of the above mentioned massive galaxies, their $r$-band absolute magnitudes, their effective radii, and their luminosities are in the ranges $-15.6 \lesssim M_r \lesssim -7.8$, 160 pc $\lesssim R_e \lesssim$ 4.1 kpc, and $0.1\times 10^6 \lesssim\left(\frac{L}{L_{\odot}}\right)_r\lesssim127 \times 10^6$, respectively. To determine whether these LSB galaxies are indeed satellites of the above mentioned massive galaxies, their distances need to be determined via further observations.}
   {Using small telescopes, we are readily able to detect LSB galaxies with similar properties to the known dwarf galaxies of the Local Group.}


  \maketitle
%

\section{Introduction}

Dwarf galaxies are the most common type of galaxies in the
universe and are crucial in testing different models of galaxy
formation and evolution \citep{gallagher94}. The properties of dwarf satellite galaxies are important for testing cosmological and gravity models on small scales \citep{Gentile07,kroupa10,Lelli15,Flores16},
and the spacial and velocity distributions of these small galaxies around their hosts probe structure formation models on scales of hundreds of kpc
\citep{kroupa05,kroupa12,kroupa15}. \citet{CK16} emphasize how different models lead to different predictions on the phase-space correlation, or lack thereof, of satellite galaxies.

In the framework of the
standard cosmological model, galaxies form in a hierarchical
process. This scenario faces challenges in providing convincing
predictions for the properties of the galaxies in the Local Group
(LG). The number of observed satellite galaxies in the LG is much
smaller than the predicted number of substructures based on numerical
simulations \citep[the ``missing satellite'' problem;][]{klypin99,moore99,bullock10,kravtsov10}. Furthermore, the majority of the most
massive predicted substructures are dense enough to trigger star
formation and hence be observable \citep[the ``too big to fail''
problem;][]{kolchin11}. These small-scale problems are often seen as  mild tensions that can be resolved with the complex processes acting amongst the constituents of the Standard Model of particle physics, i.e.  owing to baryonic physics that is not well understood (e.g. \citealt{DelPopolo14b}). Problems and successes of the standard  models based on dark matter are reviewed by \citet{DelPopolo14a}.  

In addition to these small-scale tensions, on scales spanning 10s up to hundreds of kpc,
the satellites of the Milky Way
(MW) and the Andromeda (\object{M31}) galaxies are not distributed
approximately isotropically (as expected from the standard galaxy
formation scenario) around these two massive galaxies, but instead, they
are located in thin disks and seem to have correlated orbits
\citep{kroupa05,pawlowski12,pawlowski14a,pawlowski15a,pawlowski15b,pawlowski14b,
pawlowski14c,ibata13,ibata14,ibata15}. Furthermore,
the MW and M31 satellites planes appear to be mutually correlated, and
the distribution of all non-satellite LG galaxies is highly
symmetrical in two planes equidistant from the line joining the MW and \object{M31}
\citep{pawlowski13}. And last but not least, the number of
satellites of the main galaxies of the LG (i.e. \object{M33}, MW and \object{M31}) seem
to be correlated with the masses of their bulges, amongst other
problems \citep{kroupa10, kroupa15, CK16}.

Although these features have been mostly observed and studied in the
LG, where the best spacial three-dimensional and kinematical data are available, \citet{chiboucas} have already noted the anisotropic distribution
of dwarf galaxies in the nearest galaxy group, \object{M81}, which resembles
the anisotropies evident in the LG. However, further observations are
needed to find out if LG properties are also typical  in case of other
galaxy groups. For this reason, it is vital to conduct various
systematic searches for dwarf satellite galaxies beyond the Local
Group \citep{mcconnachie12,chiboucas,sand14,spencer14,koda15,munoz15,sand15,crnojevic15,giallongo,davies,makarov}.

Recently, the potential of small-sized telescopes as tools for probing the low surface brightness (LSB) features in the nearby universe has been proposed and tested successfully. In recent years, the Stellar Tidal Stream Survey (STSS; PI. Martinez-Delgado) has obtained deep wide-field images of nearby spiral galaxies that showed evidence of being surrounded by diffuse light over-densities \citep{delgado08,delgado10,delgado15}. This observational effort has clearly revealed previously undetected stellar streams, which  makes this survey the largest sample of stellar tidal structures in the local Universe so far. Similarly, \citet{karachentsev14,karachentsev15} used small telescopes and long exposure imaging to visually search for LSB galaxies in the Local Volume. Also, the Dragonfly telephoto array of eight photographic lenses reported the discovery of seven LSB galaxies in the field of M101 \citep{merritt14}.

In this paper, we present the Dwarf Galaxy Survey with Amateur Telescopes (DGSAT), a pilot survey that exploits the deep images
obtained by the STSS during recent years to complete the census of very faint dwarf satellites around our spiral galaxy targets (that are mainly Milky Way
analogs). The first result of this project was the discovery of DGSAT I, a faint diffuse field galaxy behind the Andromeda galaxy, which turned out
to be an ultra-diffuse galaxy associated with the Pisces-Perseus supercluster \citep{delgado15b}. In this work, we present
the results of our search for faint companions around six nearby Milky Way-like galaxies of the Local Volume, namely \object{NGC 2683}, \object{NGC 3628}, \object{NGC 4594} (\object{M104}), \object{NGC 4631}, \object{NGC 5457} (\object{M101}), and \object{NGC 7814}. Throughout this work, the adopted distances and distance moduli for these galaxies are the mean values given by the NASA/IPAC Extragalactic Database (NED)\footnote{The NED is operated by the Jet Propulsion Laboratory, California Institute of Technology, under contract with the National Aeronautics and Space Administration.}.

This manuscript is organized as follows. The observations are explained in Section \ref{sec:obs}. We describe our automated data calibration in Section \ref{sec:data} and our searching strategy and parameter extraction in Section \ref{sec:analysis} . Our results for each individual field are presented in Section \ref{sec:results} and discussed in Section \ref{sec:discussion}. Finally, we summarize our results is Section \ref{sec:conclusion}.


\section{Observations}\label{sec:obs}
\begin{figure}
\includegraphics[scale=0.35]{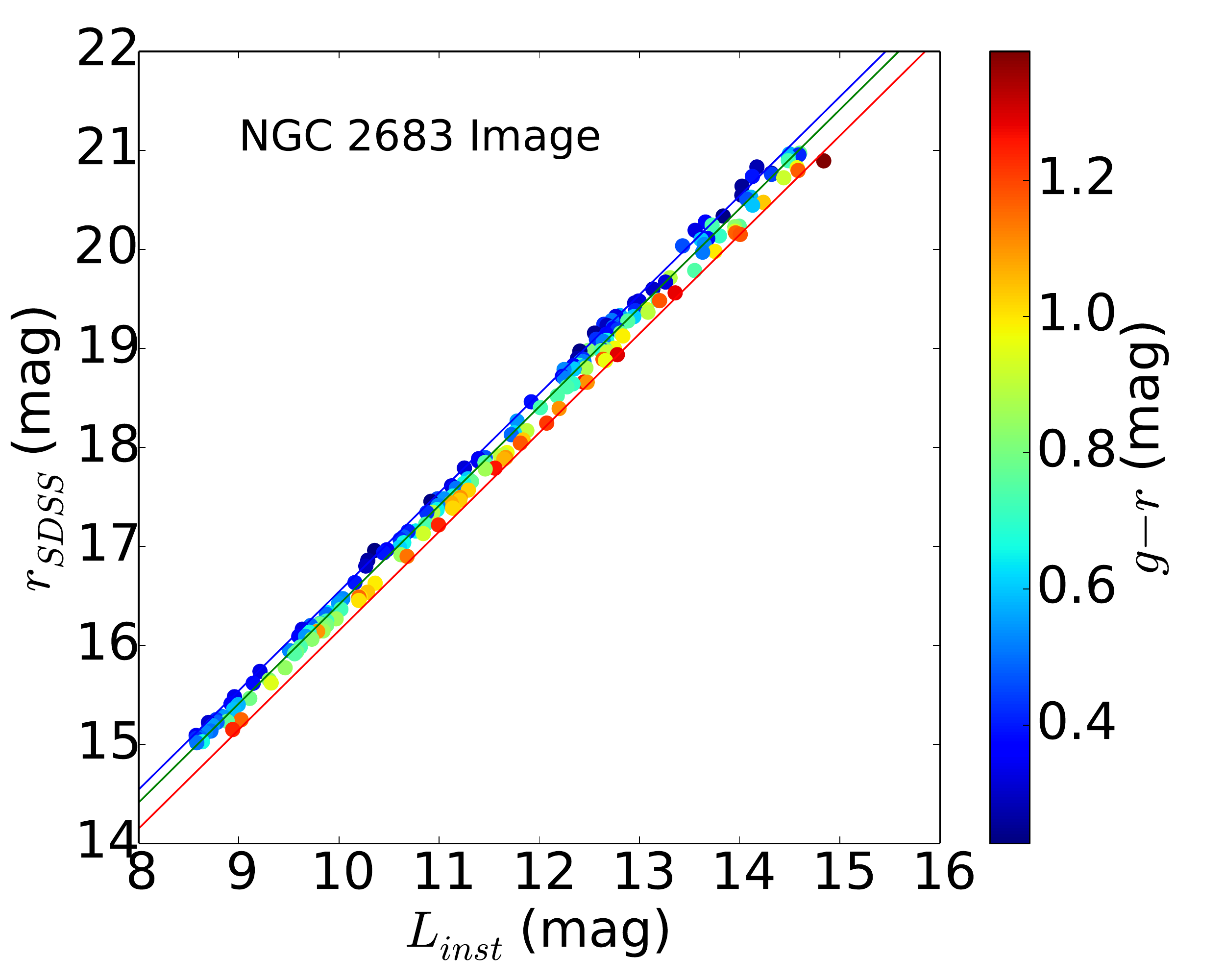}
\includegraphics[scale=0.35]{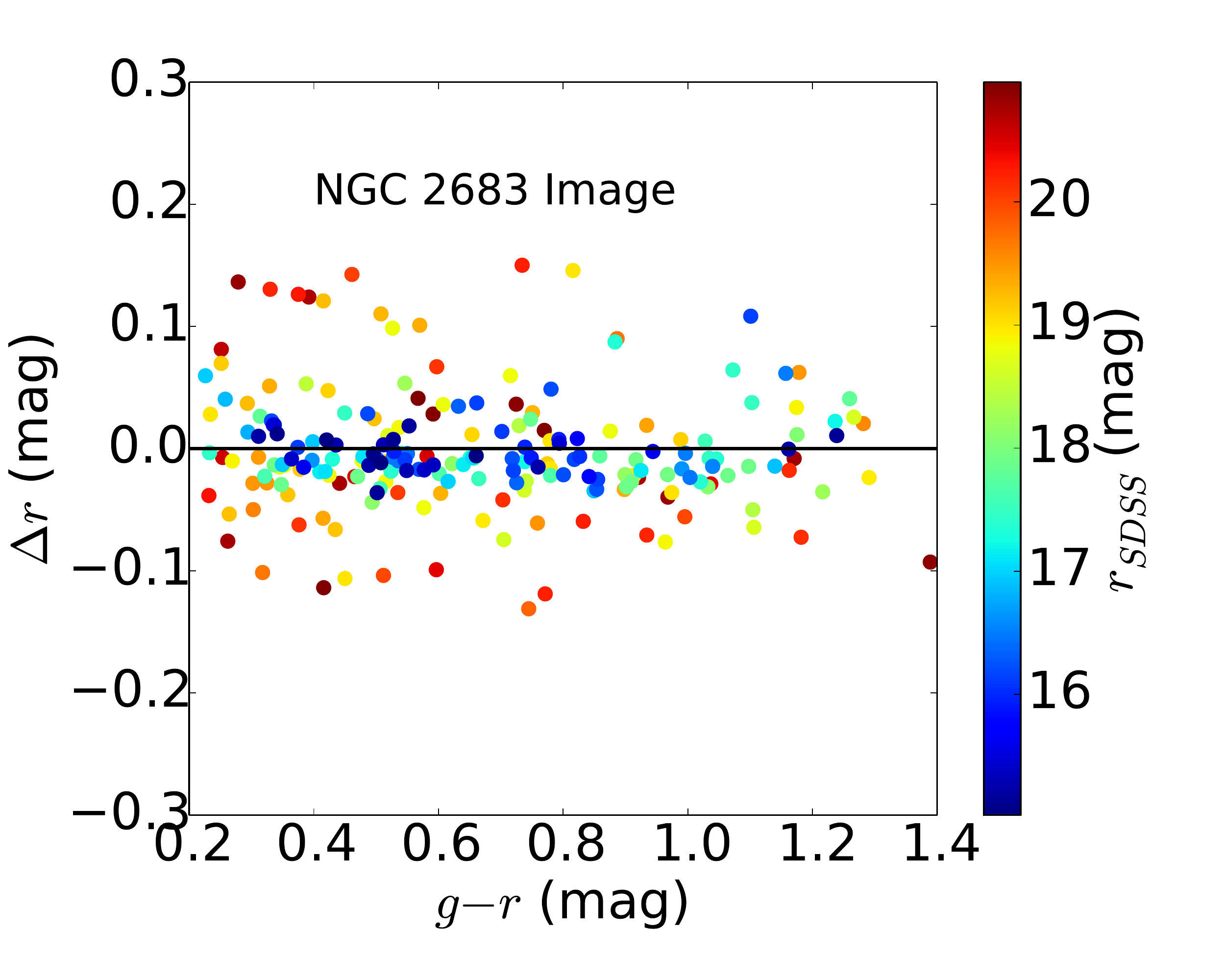}
\caption{An example of the results of calibration to the SDSS $r$-band for the \object{NGC 2683} field. Top: SDSS $r$ magnitude $r_{SDSS}$ of the calibrating stars in the image compared to the corresponding instrumental measurement $L_{inst}$ ;  the colour code is $(g-r)$. The red, green, and  blue solid lines are the $r_{cal}$ from Eq. \ref{eq:rcal} evaluated at the maximum (reddest),  median, and minimum (bluest) of the $(g-r)$ values of the sample, respectively. Bottom: magnitude residuals after calibration, $\Delta r = r_{SDSS}-r_{cal}$, vs. $(g-r);$  the colour code is $r_{SDSS}$. No dependency on $(g-r)$ remains. The standard deviation of the residuals in this particular case is 0.05 mag and we reach similar accuracy in calibration of the rest of images (see Table \ref{t:galaxies}). \label{fig:calib}}
\end{figure}

Detecting faint features in the halos of spiral galaxies requires wide-field, deep images taken in very dark sky conditions and with exquisite flat-field quality over a wide region ($>$ 30 arcmin) around the targets. Our survey strategy strives for multiple deep exposures of each target using high throughput clear filters with near-IR cut-off, known as luminance ($L$) filters \citep[4000 \AA $< \lambda <$7000 \AA; see Fig. 1 in][]{delgado15} and a typical exposure time of 7-8 hours. Our typical 5$\sigma$ surface brightness detection limit (measured in random $\sim$3 arcsec diameter apertures) is $\sim$ 28.5 and 28 mag/arcsec$^2$ in $g$ and $r$ bands, which is approximately three magnitudes deeper than the Sloan Digital Sky Survey (SDSS) II images \citep{delgado10}. From a direct comparison with SDSS data, we find that our images are ten times deeper than this survey in terms of photon statistics with comparable systematic background uncertainties \citep{delgado10,cooper11}.\\

The observations of DGSAT were conducted with a network of privately owned robotic observatories equipped with modest-sized telescopes (0.1-0.8 meter) located in Europe, the United States, Australia, and Chile. Each observing location features spectacularly dark, clear skies with seeing below 1.5$''$. The list  of participating facilities that have provided data for this paper is given in Table \ref{t:telescopes}. Each telescope is equipped with a latest generation astronomical commercial CCD camera, using in the majority of the cases a film-format, 16 mega pixel imaging sensor composed of a 4096 $\times$ 4096 pixel array with 9 $\times$ 9 micron pixels. The fields of view and pixel scales resulting from the telescope and camera combinations are listed in Table 1. Additionally, our survey uses portable apocromatic refractors (with a typical aperture of 0.1 - 0.15 m) for achieving wider fields around those nearby galaxies with an extended apparent size (e.g. M101). The data reduction follows standard techniques (zero level subtraction and first order flat-field corrections from a large number of  dome and twilight flat frames) using pipelines successfully proven during the STSS \citep{delgado08,delgado10,delgado15}.

\begin{table*}
 \begin{center}
  \caption{Target galaxies in order of NGC numbers, observatory, location,  telescope and its focal ratio f/ used for observing, observation dates, total exposure time, t$_{exp}$ (in seconds), field of view (FOV) (in arcmin$^2$), and pixel scale (in arcsec) of the images. \label{t:telescopes}}
  
  \begin{tabular}{lcccccccc}
Target& Observatory & Location & Telescope & f/ & Obs. Date & t$_{exp}$ (s) & \begin{tabular}{@{}c@{}}FOV \\ (arcmin$^2$)\end{tabular} &  \begin{tabular}{@{}c@{}}pixel scale \\ (arcsec/pix)\end{tabular}\\
  \hline
\object{NGC 2683}  & {\small ROSA(POLLUX)  } &  {\small   Verclause, France  }&   {\small    Newton 0.4-m }&  3.5  &  Feb-Mar 2015   &      24000    &      81$\times$81  &  1.22\\
\object{NGC 3628}  & {\small DGRO-Rancho Hidalgo} &{\small New Mexico, USA  }&    {\small     RCOS 0.36-m }&   7.9  &  Dec 2011   &           36000  &        43$\times$43  &   0.62 \\
\object{NGC 4594}  &{\small Riverdingo }&       {\small  Adelaide, Australia }&   {\small   RCOS 0.36-m }&   7.9 &    Apr 2009     &  25200  &         43$\times$ 43          & 0.62\\
\object{NGC 4631}  & {\small Black Bird } &     {\small   New Mexico, USA }&    {\small RCOS 0.5-m    }& 8.1 &      May-Nov 2011   & 63000& 31$\times$ 31 & 0.43\\           
\object{NGC 5457}  & {\small ROSA(POLLUX)} &    {\small    Verclause, France }&  {\small       Newton 0.4-m }&   3.5  &  Apr-May  2014&         75600  &        81$\times$81 &   1.22 \\
\object{NGC 5457}  & {\small Antares } &       {\small Gossau, Switzerland }&    {\small     TEC140 APO }&   7.2  &  Feb - May 2012&         87600  &        121$\times$80 &   1.82 \\
\object{NGC 7814}  &{\small Rancho del Sol} &   {\small   California, USA }&     {\small      RCOS 0.5-m }&    8.3   &       Aug 2013       &         40800 &   29$\times$29       &          0.43\\
  \hline
  \end{tabular}
 \end{center}
\end{table*}

%
\section{Data calibration}\label{sec:data}
We developed a semi-automatic pipeline for calibrating our flux measurements to the SDSS $r$-band, as explained below. Since the applied Luminance filter, $L$, is a wide band filter that almost covers the entire SDSS $g$ and $r$ bands, a colour dependency is expected between the instrumental magnitude measurement, $L_{inst}$, and $r$ magnitude. Therefore, the calibration takes the form

\begin{equation}
\label{eq:rcal}
 r_{cal}=c_0L_{inst}+c_1(g-r)+c_2
,\end{equation}
where $r_{cal}$ is the calibrated $r$ magnitude, $c_0$ tunes the linear relation between $r$ and $L_{inst}$, $c_1$ corrects the colour dependency, and $c_2$ is just a magnitude zero-point correction. These calibrating parameters are determined via taking a maximum likelihood approach and minimising the function

\begin{equation}
\label{eq:chi}
 \chi^2(c_j)=\sum_i^{N_{star}} \frac{[c_0L_{inst,i}+c_1(g-r)_i+c_2-r_i]^2}{\sigma^2_{L_{inst,i}}},
\end{equation}
where $\sigma_{L_{inst}}$ is the uncertainty of the instrumental flux measurement and $N_{star}$ is the number of calibrating stars. To minimize the effect of human bias, we do not select ``good stars'' for calibrating, but those that are used for the final calibration are selected automatically via the following approach. In our pipeline, first the SExtractor \citep{sextractor} is run on the image (or a portion of it, depending on the field of view and the number of available stars) to detect and measure the flux of all the objects in the image. After that, the coordinates of our detected objects are cross-identified with the SDSS DR12 catalogue\footnote{\url{http://skyserver.sdss.org/dr12/en/tools/crossid/crossid.aspx}} and only the stars with magnitude $r \geq 15$ pass to the next step. The reason for this magnitude cut is that \citet{chonis08} found that the SDSS photometry suffers from saturation effects around $g,r,i \approx 14$ mag. In the next step, the stars outside the colour ranges $0.08<(r-i)<0.5$ and $0.2<(g-r)<1.4$ are rejected since they behave non-linearly in the $(r-i)$ vs. $(g-r)$ space \citep[see][]{chonis08}. After this, we have a large number of stars (different for various images) with $g$ and $r$ magnitudes from SDSS and the $L_{inst}$ from FLUX-AUTO given by the SExtractor. Using these stars, our code minimizes the $\chi^2$ in eq. (\ref{eq:chi}) for the first estimation of the $c_j$ with which $r_{cal}$ (from eq. \ref{eq:rcal}) and $\Delta r = r_{SDSS}-r_{cal}$ are computed. At the next step, the $3\sigma$ outliers from the best-fit result (i.e. the stars with $\Delta r > 3\sigma$) are rejected, where $\sigma$ is the standard deviation of $\Delta r$. The fitting is repeated with the remaining stars and is followed by another outlier rejection as explained above. This procedure is repeated until the fitting reaches a point where no $3\sigma$ outliers remain. Then the final $c_j$ are obtained and the final $r_{cal}$ and $\Delta r$ are determined. An example of the results of our calibration for the image of the field of \object{NGC 2683} is shown in Figure \ref{fig:calib}. The top panel shows the SDSS $r$ magnitudes of the calibrating stars vs. their instrumental measurement $L_{inst}$ and the bottom panel illustrates that the final magnitude residuals after calibration, $\Delta r$, show no dependency on $(g-r)$ colour. The latter means that our colour correction has well accounted for the systematic effects due to the spectral energy distribution of measured objects and the width difference of the transmitting filters. The final number of stars used for the calibration, $N_{star}$, and the standard deviation of the final residuals, $\sigma_{cal}$, (which is added in quadrature to the subsequently reported results) are listed in Table \ref{t:galaxies} for all the images. If  one chooses the calibrating stars manually instead of the above explained
automatic star selection, a smaller but not necessarily reliable $\sigma_{cal}$ might be achieved. The method we present here is statistically robust.

\begin{figure*}
\begin{center}
 \includegraphics[scale=0.5]{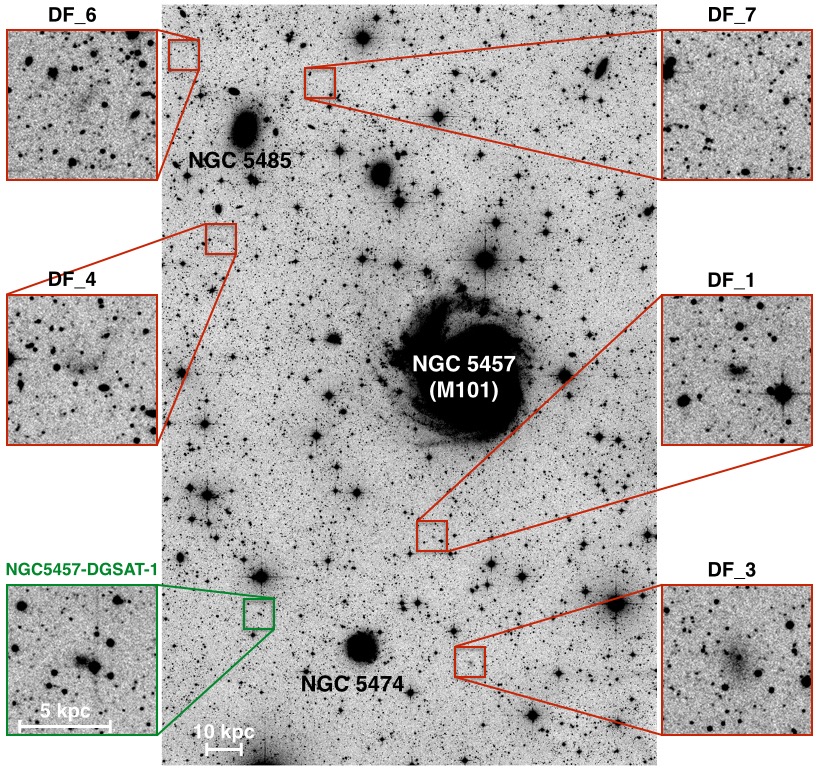}
 \caption{The $2^\circ\!\!.0 \times 1^\circ\!\!.3$ field of \object{NGC 5457} (\object{M101}) or
Pinwheel galaxy. North is up and east is to the left. The zoomed-in squares show the low surface brightness galaxies in the image. The red squares are five dwarf galaxies reported by the Dragonfly project (DF-number), which are all confirmed. The green square is the one newly discovered in our image (\object{NGC5457-DGSAT-1}). Two of the Dragonfly LSB galaxies, namely the DF-2 and DF-5, are outside the field of our image.\label{fig:m101}}
\end{center}
\end{figure*}
The limiting surface brightnesses of our images were determined following the method described in \citet{delgado10}. For estimating the photon noise, we measured the standard deviation of apertures with around 3 arcsec diameter and computed the surface brightness corresponding to five times that standard deviation. The obtained values for each image are listed in Table \ref{t:galaxies}.

\begin{table*}
 \begin{center}
  \caption{Target galaxies in order of NGC numbers, their adopted distances, d, and distance moduli, DM, from the NED, number of final calibrating stars, $N_{star}$, and final standard deviation in $\Delta r$, $\sigma_{cal}$, and the 5 $\sigma$ values of the limiting surface brightness (photon noise) in r band for each image in mag/arcsec$^2$ using boxes of around 3 arcsec.\label{t:galaxies}}
  
  \begin{tabular}{lccccc}
  Target & d (Mpc)& DM (mag) & $N_{star}$ & $\sigma_{cal}$ (mag) & \begin{tabular}{@{}c@{}}Photon Noise \\ (mag/arcsec$^2$)\end{tabular}\\
  \hline
  \object{NGC 2683} & $10.5 \pm 2.2$ & $30.05\pm0.50$ &212 & 0.05 & 28.03 \\
  \object{NGC 3628} & $10.9 \pm 2.4$ & $30.14\pm0.53$ &138 & 0.07 & 28.06   \\
  \object{NGC 4594} & $11.1 \pm 4.1$ & $30.12\pm0.69$ &140 & 0.08 & 27.54 \\
  \object{NGC 4631} & $5.8 \pm 1.5$ & $28.76\pm0.58$ &72 & 0.08 & 29.61  \\
  \object{NGC 5457} & $6.8 \pm 0.8$ & $29.16\pm0.26$ &104 & 0.05 & 28.92  \\
  \object{NGC 7814} & $16.4 \pm 3.6$ & $31.03\pm0.49$ &107 & 0.05 & 27.87 \\
  \hline
  \end{tabular}
 \end{center}
\end{table*}

\begin{figure*}
\begin{center}
 \includegraphics[scale=0.5]{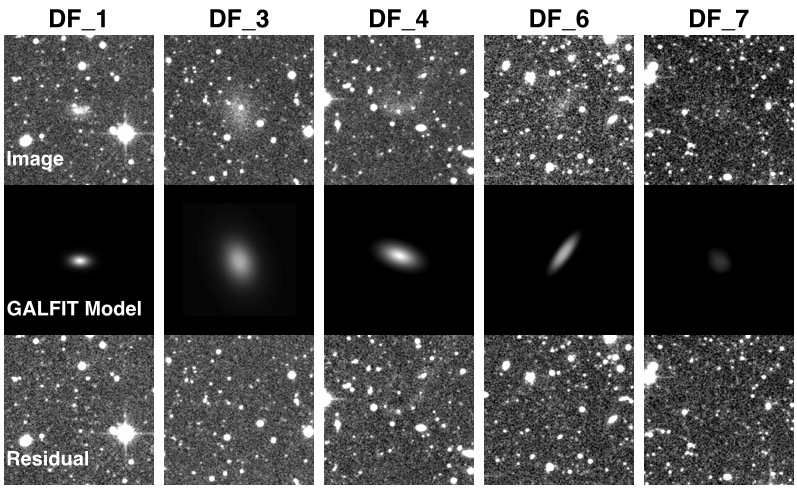}
 \caption{Top:  Dragonfly LSB galaxies (DF-number) detected in our image. DF-2 and DF-5 are outside our field. Middle: our GALFIT models (see Table \ref{t:results} for the results of the modelling). Bottom:  residual images obtained by subtracting the model image (middle) from the main image (top). The results are in agreement with those reported by the Dragonfly team (see Figure \ref{fig:comparison}).\label{fig:m101_galfit}}

\end{center}
\end{figure*}

\section{Methods of analysis}\label{sec:analysis}
In this section we explain our searching strategy to find dwarf galaxy candidates and to extract information about their structural properties by light profile fitting. The following two subsections outline our general procedures, while we go into detail  later when we discuss individual galaxies.

\subsection{Searching strategy}\label{subsec:searching}
To search for low surface brightness galaxies, we use the SExtractor software by \citet{sextractor}. Though the top-hat convolution filter in this software is suited for detecting extended low surface brightness objects, we do not limit the search only to that filter and run SExtractor several times with different convolution filters and detection thresholds. The detected objects from different runs are matched based on their image coordinates. This gives us a large catalogue of all the objects in the image. We put a size constraint on the detections and remove all the objects with a FLUX-RADIUS smaller than 1.5 times the size of the full width at half maximum (FWHM) of the point spread function (PSF) of the image. This value (which is between 4 to 6 arcsec for our images) seems to be large enough for rejection of most of the stars and small enough to keep the smallest LSB galaxies. Most of the rejected objects are stars and the rest are bright background galaxies or noise. This cut condition helps to  reduce significantly the number of detections that need to be checked visually. Among the surviving detections, most of them are either noise fluctuation, saturated stars and their wings, or galactic cirrus.

\begin{figure}
 \begin{center}
 \includegraphics[scale=0.27]{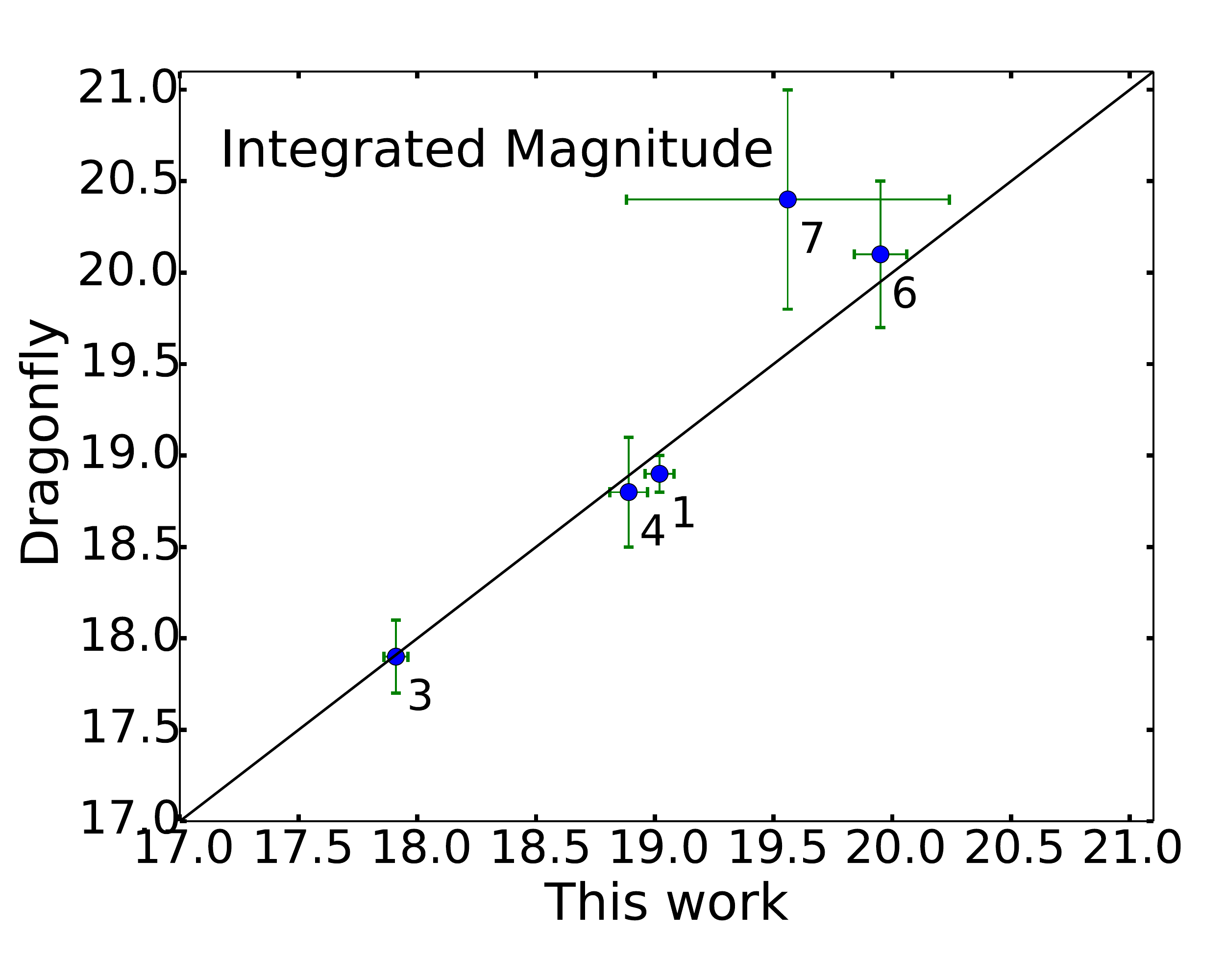}
 \includegraphics[scale=0.27]{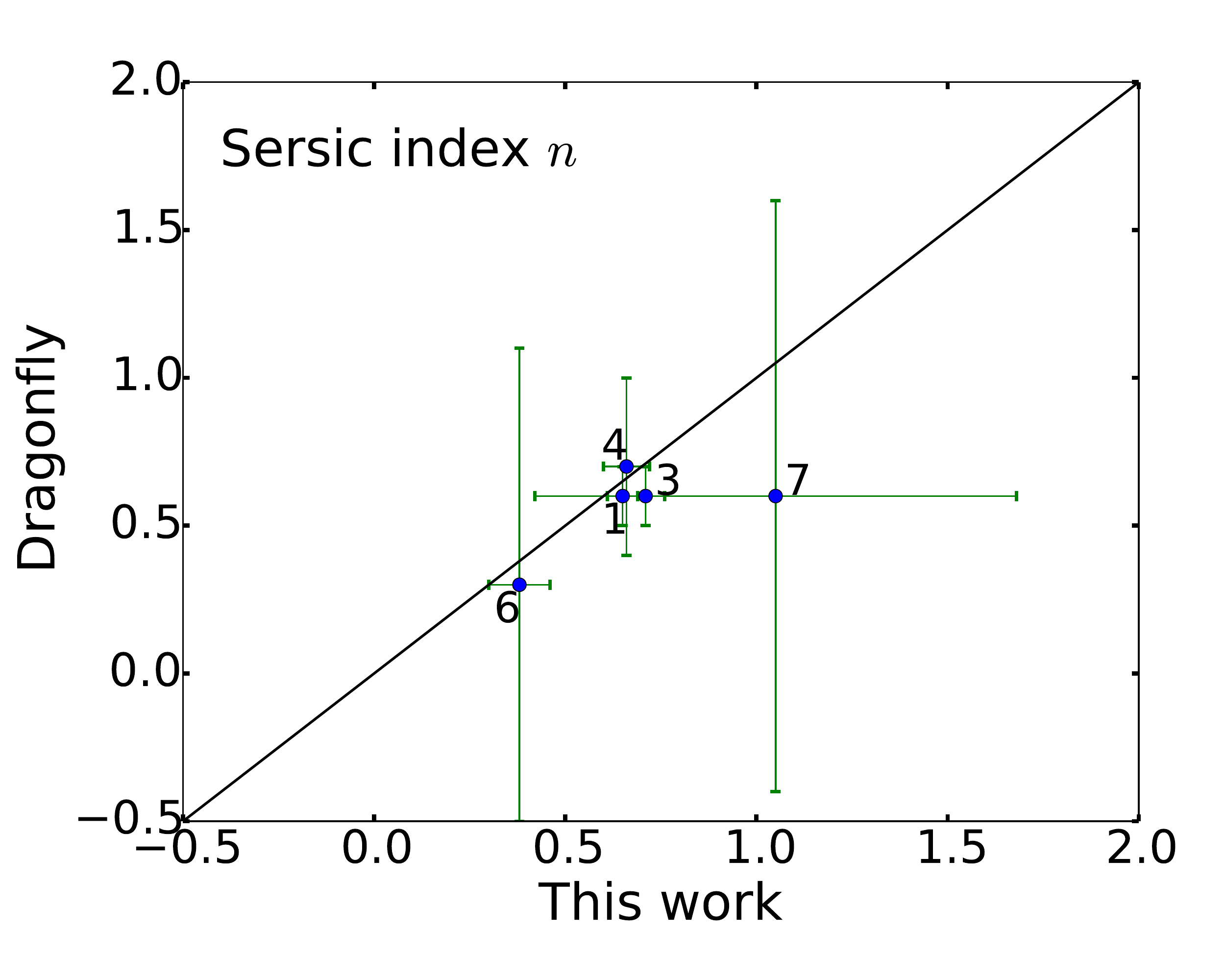}
 \includegraphics[scale=0.27]{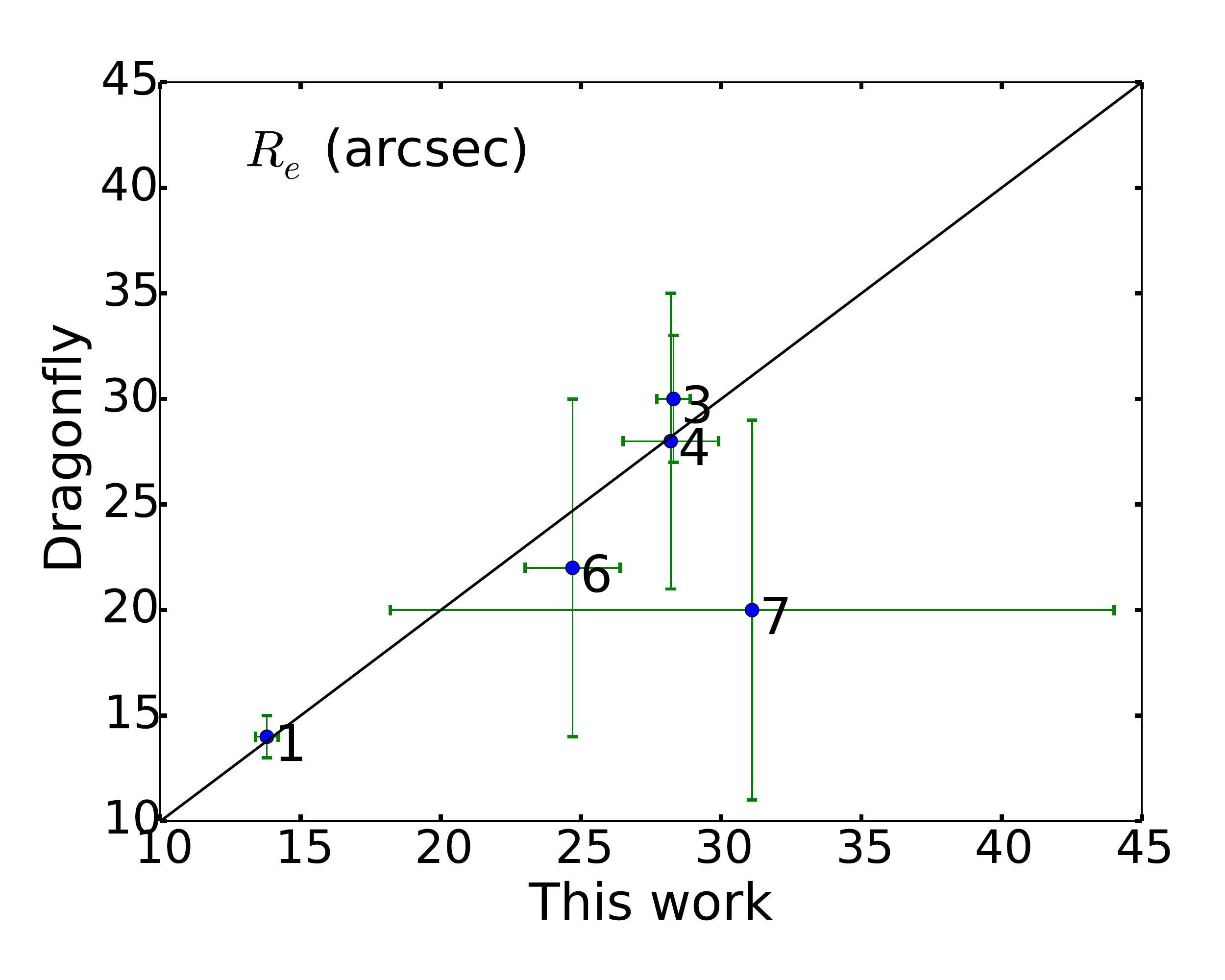}
  \caption{Comparison between our results for integrated magnitude (in $g$ band, assuming $(g-r)=0.5$), Sersic index $n$, and effective radius $R_e$ with those of the Dragonfly study \citep{merritt14} for DF1, 3, 4, 6, and 7.\label{fig:comparison}}
 \end{center}
\end{figure}

\begin{figure}
\begin{center}
 \includegraphics[scale=0.4]{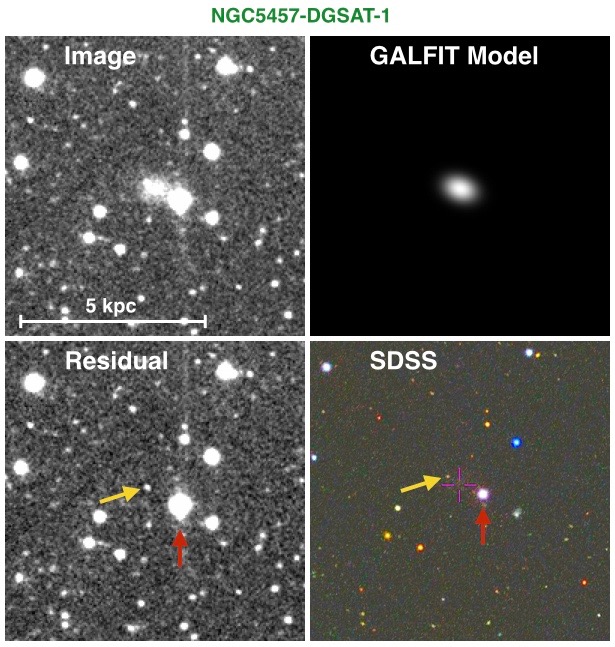}
 \caption{Top left: cut out portion of our image centred on NGC5457-DGSAT-1. Top right: our GALFIT model (see Table \ref{t:results} for the results of the modelling). Bottom left: residual image obtained by subtracting the model image from the main image. Bottom right: same field from SDSS for comparison. In both of the bottom images, the red arrow points to USNO-A2.0 1425-08068454 and the yellow arrow points to an uncatalogued point source. The comparison of these two images with the main image (top left) helps to better see NGC5457-DGSAT-1.\label{fig:dgsat1}}

\end{center}
\end{figure}

\begin{figure}
\begin{center}
 \includegraphics[scale=0.3]{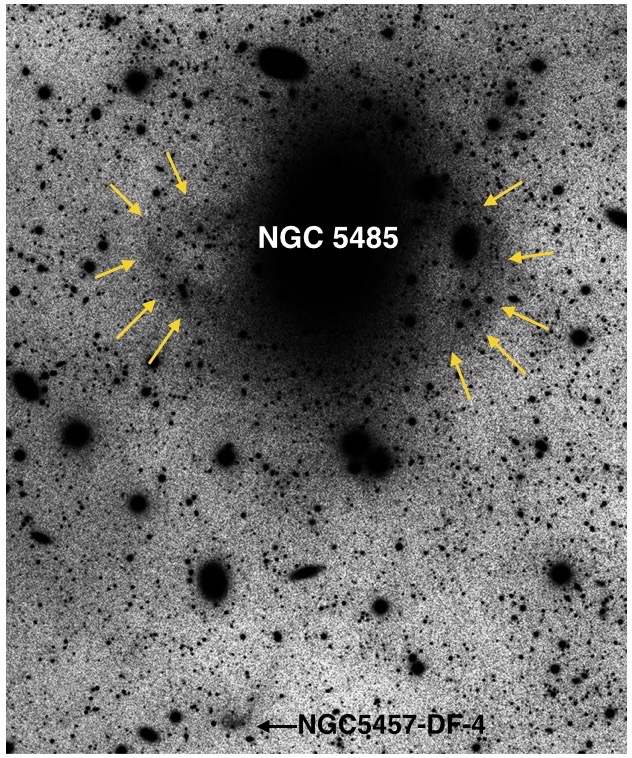}
 \caption{Low surface brightness stellar tidal stream around the elliptical galaxy \object{NGC 5485} is shown by yellow arrows. North is up and east is to the left. \object{NGC5457-DF-4} is also indicated by name (see Figure \ref{fig:m101}).\label{fig:5485}}

\end{center}
\end{figure}

\subsection{Parameter extraction}\label{subsec:params}
To determine the structural properties of the dwarf candidates, we fit a Sersic model \citep{sersic} on them via the GALFIT software developed by \citet{galfit}. For this purpose, we cut a proper-sized square centred on each dwarf candidate. Depending on the brightness and position of possible neighbouring or overlapping stars, we either mask them, remove them by PSF photometry via IRAF/DAOPHOT software \citep{daophot} before model fitting, or fit them simultaneously with the dwarf candidate using GALFIT. We repeat the fitting until we reach a good fit by checking the reduced $\chi^2$ (given by GALFIT) and by visually inspecting the model and residual images.

\section{Results}\label{sec:results}
In this section we report the results of our search for dwarf satellite galaxies. We start with the field of the M101 galaxy, which has the largest number of newly detected dwarf galaxies that have been recently reported by the Dragonfly team \citep{merritt14}. Therefore, this galaxy is the proper target to begin with. The other fields, where most of our new detections are located, are reported in order of NGC number.

\subsection{NGC 5457 (M101) revisited}\label{subsec:m101}
\object{M101} or the Pinwheel galaxy is a massive spiral galaxy at an adopted distance of $6.8 \pm 0.8$ Mpc (Table \ref{t:galaxies}). The field of M101 has been recently probed by the Dragonfly project \citep{merritt14} for low surface brightness features. They had a $3^\circ\!\!.3 \times 2^\circ\!\!.8$ image centred on M101 and were able to discover seven low surface brightness galaxies. We use our $2^\circ\!\!.0 \times 1^\circ\!\!.3$ image (Figure \ref{fig:m101}) of the field of this galaxy (which is a combination of two images taken by K. Teuwen and F. Neyer) to revisit the discovered Dragonfly LSBs and look for other possible dwarf galaxy candidates. As they mention in \citet{merritt14}, their survey is not able to reveal the smallest galaxies because of their relatively large FWHM of around 6.5 arcsec. Our image of the M101 field with an FWHM $\approx3.2$ arcsec (corresponding to around 100 pc at the distance of M101) is more suitable than that of the Dragonfly for revealing small galaxies.

Two of the M101 LSB galaxies discovered by Dragonfly, namely DF2 and DF5, are outside of our image field. With our search strategy explained in Section \ref{subsec:searching}, we could not only detect \object{DF1}, \object{DF3}, \object{DF4}, \object{DF6} and \object{DF7}, but we also find yet another low surface brightness galaxy (the \object{NGC5457-DGSAT-1}) which was not reported by the Dragonfly project. This dwarf galaxy candidate was also independently discovered by \citet{karachentsev15}\footnote{The image of M101 used in our study, taken by the astrophotographer F. Neyer (see Table \ref{t:telescopes}), is also used in that paper.}. The Dragonfly LSB galaxies and our obtained models for them are shown in Figure \ref{fig:m101_galfit}. The top panel is the cut-out portion of the main image centred on each LSB galaxy, the middle panel shows the model obtained by GALFIT and the lower panel is the result of the subtraction of the model from the main image. Prior to the GALFIT modelling, the foreground stars were removed by fitting the PSF model of the image using DAOPHOT. The models can be compared to those reported in Figure 1 of \citet{merritt14}. For almost all of them, the residual images are clean of the LSB galaxy but in the case of DF4 there remains a bit of unsubtracted light. This galaxy has an attached, elongated tail , which is not visible in the Dragonfly image, and may be caused by tidal disruption of this galaxy. The results of the modelling for DF galaxies are shown in the top part of Table \ref{t:results}. We assume $(g-r)=0.5,$ which is the mean value for these galaxies measured by \citet{merritt14}\footnote{We assume the same value for other dwarf galaxy candidates that we present later in this paper. Assuming that the dwarf galaxies are composed by metal-poor, old populations, we can expect an integrated colour similar to those of Galactic globular clusters \citep{vanderbeke}.}. Figure \ref{fig:comparison} gives a comparison of our results with those of the Dragonfly team. This agreement further confirms the consistency of our magnitude calibration and galaxy-modelling approach.

The newly discovered galaxy in the field of M101 and its model are shown in Figure \ref{fig:dgsat1}. We also included the SDSS image of the same field for comparison. In the case of \object{NGC5457-DGSAT-1}, the bright neighbouring star was masked before fitting a Sersic model to the galaxy. In the resulting residual image, the bright star (\object{USNO-A2.0 1425-08068454}, indicated with a red arrow) and the uncatalogued point source (identified by a yellow arrow) can be easily compared with those in the SDSS image. The results of the modelling for \object{NGC5457-DGSAT-1} are also shown in Table \ref{t:results}. The reason that Dragonfly could not detect this object is most likely because their FWHM is larger and, therefore, this LSB galaxy and its neighbouring stars were not resolved properly. Another interesting feature in our image of the field of M101 is illustrated in Figure \ref{fig:5485}. It shows a clear stellar tidal stream around the background elliptical galaxy \object{NGC 5485}, which is situated at a distance of around 28 Mpc. This ring-like LSB feature was also noted by \citet{karachentsev15} and could be related to the tidal disruption of a dwarf satellite with an almost circular orbit.


\begin{figure*}
 \begin{center}
  \includegraphics[scale=0.5]{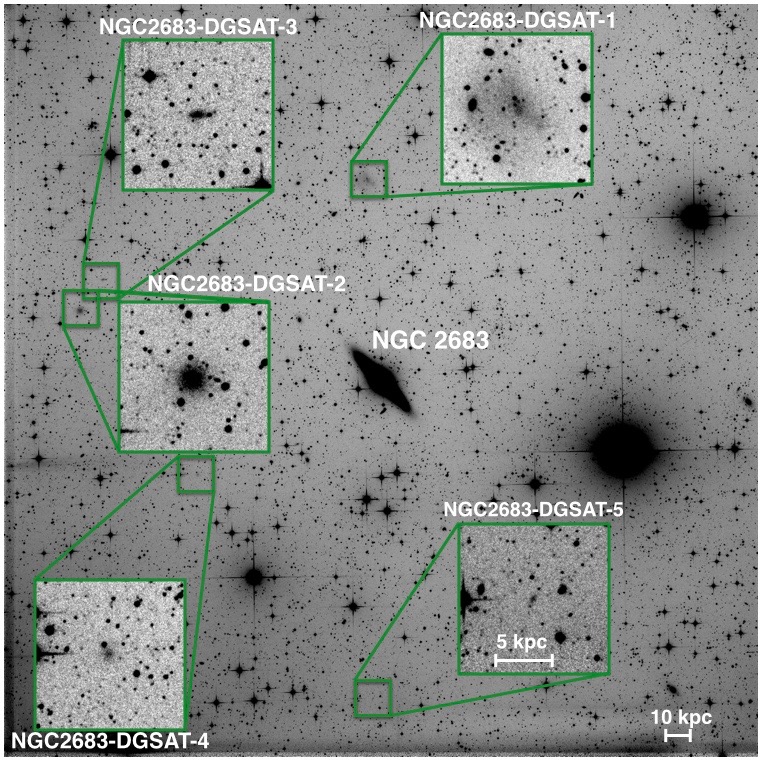}
  \caption{The $1^\circ\!\!.4 \times 1^\circ\!\!.4$ field of \object{NGC 2683}. North is up and east is to the left. The zoomed-in squares show the LSB galaxies in the image.\label{fig:ngc2683}}
  
 \end{center}
\end{figure*}

\begin{figure*}
 \begin{center}
  \includegraphics[scale=0.35]{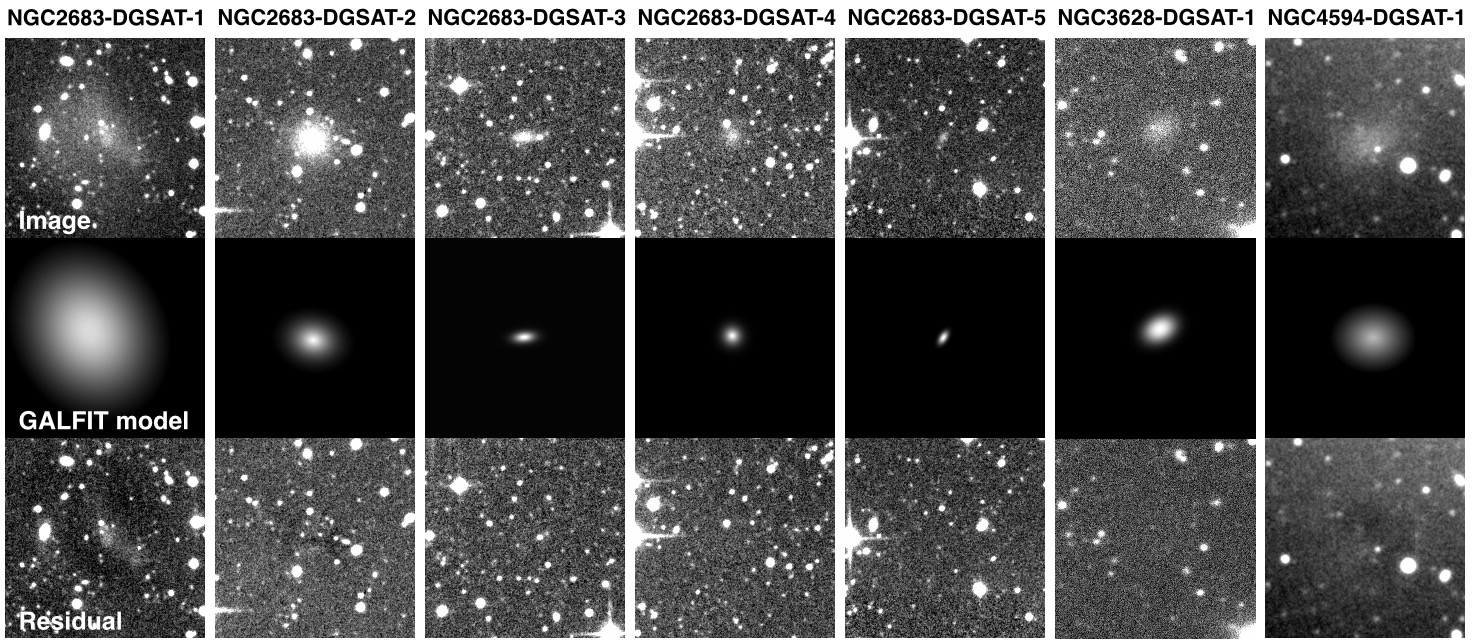}
  \includegraphics[scale=0.35]{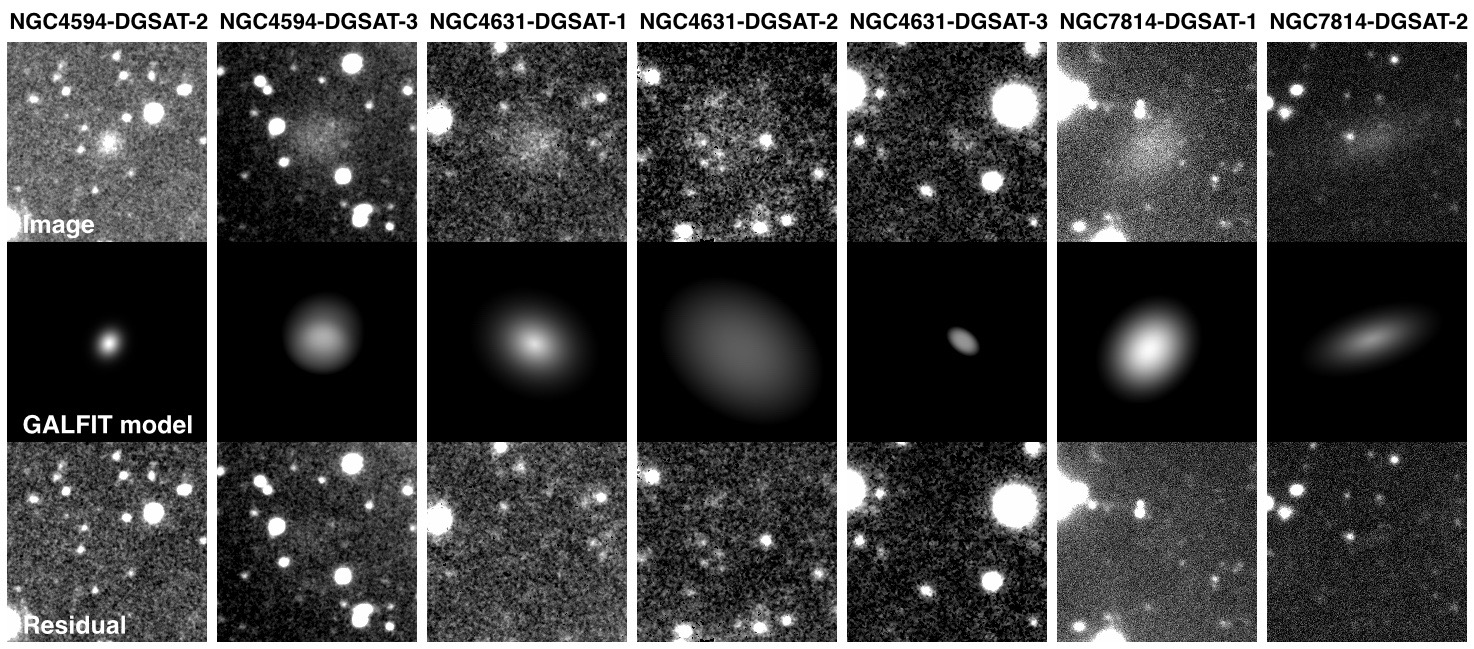}
  \caption{In both the top and bottom panels, the first row: dwarf galaxy candidates detected in our images (except that of the NGC 5457); the middle row: our GALFIT models (see Table \ref{t:results} for the results of the modelling), and the bottom row:  residual images obtained by subtracting the model image (middle) from the main image (top).\label{fig:dgsats}}
  
 \end{center}
\end{figure*}

\subsection{The NGC 2683 field}\label{subsec:2683}
\object{NGC 2683} is a galaxy with a bulge and an active nucleus at an adopted distance of $10.5 \pm 2.2$ Mpc (Table \ref{t:galaxies}). Our image of the field of this galaxy (Figure \ref{fig:ngc2683}) has the dimensions of $1^\circ\!\!.4 \times 1^\circ\!\!.4$, a pixel scale of 1.22 arcsec/pixel, and its FWHM is $\approx3.5$ arcsec. We zoomed into five LSB galaxies in this image. The two largest and brightest of these LSB galaxies were already catalogued in \citet{karachentsev04} as [KK98a] 084944.1+335913 and [KK98a] 085216.3+334502. We refer to these galaxies as \object{NGC2683-DGSAT-1} and 2 in our catalogue. The former has an irregular shape, while the latter is almost spheroidal. Similar to the other LSB galaxies in this paper, we fit a Sersic model to these two galaxies. The results are shown in Figure \ref{fig:dgsats} and Table \ref{t:results}. In the case of NGC2683-DGSAT-1, because of its irregular shape and many foreground stars on top of its image, it is very difficult to obtain an accurate model. Therefore, its GALFIT model and the corresponding measurements in this paper are only estimates. As can be seen in the figure, we could also discover three other dwarf galaxy candidates in the field of \object{NGC 2683}. During the final phase of the writing of this paper, we noted that one of these dwarf galaxy candidates, namely \object{NGC2683-DGSAT-4}, was independently discovered by \citet{karachentsev15}. Results of modelling \object{NGC2683-DGSAT-3} to 5 can also be seen in Figure \ref{fig:dgsats} and Table \ref{t:results}.

\begin{table*}
 \begin{center}
  \caption{LSB galaxies, their equatorial coordinates, RA and DEC, and their structural parameters evaluated by GALFIT modelling; calibrated integrated magnitude, $r_{cal}$, surface brightness, $\mu_{e}$, at effective radius, the Sersic index, $n$, the effective radius, $R_e$, and the axis ratio, $b/a$. In the case of NGC5457-DF-7, the axis ratio is kept fixed (to the value given by SExtractor) to stabilize the fitting. \label{t:results}}
  
  \begin{tabular}{llllllll}
  
 ID & RA & DEC & $r_{cal}$ {\scriptsize(mag)} & $\mu_{e}$ {\scriptsize(mag.arcsec$^{-2}$)} & $n$ & $R_e$ {\scriptsize (arcsec)} & $b/a$\\
  \hline
 {\small \object{NGC5457-DF-1}} & 14:03:45.0 & +53:56:38.0  & $18.52\pm0.06$  & $26.14\pm0.06$  & $0.65\pm0.04$  & $13.8\pm0.4$  & $0.59\pm0.01$ \\
 {\small \object{NGC5457-DF-3}} & 14:03:05.8 & +53:36:52.0  & $17.41\pm0.05$  & $26.58\pm0.05$  & $0.71\pm0.05$  & $28.3\pm0.6$  &  $0.56\pm0.01$ \\
 {\small \object{NGC5457-DF-4}} & 14:07:33.8 & +54:42:39.2  & $18.39\pm0.08$  & $27.46\pm0.07$  & $0.66\pm0.06$  & $28.2\pm1.7$   & $0.53\pm0.02$  \\
 {\small \object{NGC5457-DF-6}} & 14:08:18.7 &  +55:11:30.6 & $19.45\pm0.11$   & $27.43\pm0.11$  & $0.38\pm0.08$  & $24.7\pm1.7$  &  $0.31\pm0.02$  \\
 {\small \object{NGC5457-DF-7}} & 14:05:47.5 & +55:07:57.3  & $19.06\pm0.68$  & $28.85\pm0.68$   & $1.05\pm0.63$  & $31.1\pm12.9$  & $0.70$(fixed)   \\
{\small \object{NGC5457-DGSAT-1}} & 14:06:49.8 & +53:44:28.8 &$18.49\pm0.10$  & $25.69\pm0.10$ & $0.47\pm0.07$ & $10.8\pm0.9$ & $0.72\pm0.06$\\
\hline
{\small \object{NGC2683-DGSAT-1}} & 8:52:47.8 & +33:47:33.1 & $14.49\pm0.07$ & $26.29\pm0.07$ & $0.61\pm0.05$ & $80.9\pm5.8$ & $0.82\pm0.01$ \\
{\small \object{NGC2683-DGSAT-2}} & 8:55:23.3 & +33:33:32.4 & $16.11\pm0.06$ & $25.62\pm0.06$ & $0.78\pm0.02$ & $27.3\pm0.7$ & $0.79\pm0.01$ \\
{\small \object{NGC2683-DGSAT-3}} & 8:55:10.9 & +33:36:45.7 & $18.59\pm0.07$ & $25.52\pm0.07$ & $0.61\pm0.04$ & $11.2\pm0.1$ & $0.61\pm0.02$ \\
{\small \object{NGC2683-DGSAT-4}} & 8:54:20.0 & +33:14:49.1 & $18.99\pm0.23$ & $26.48\pm0.23$ & $0.70\pm0.19$ & $9.9\pm2.1$ & $0.98\pm0.14$ \\
{\small \object{NGC2683-DGSAT-5}} & 8:52:48.1 & +32:49:37.7 & $20.08\pm0.08$ & $26.00\pm0.08$ & $0.52\pm0.05$ & $6.8\pm0.4$ & $0.55\pm0.03$ \\
\hline
{\small \object{NGC3628-DGSAT-1}} & 11:21:37.0 & +13:26:50.7 & $19.21\pm0.08$ & $25.78\pm0.08$ & $0.46\pm0.03$ & $8.2\pm0.3$ & $0.74\pm0.02$ \\
\hline
{\small \object{NGC4594-DGSAT-1}} & 12:39:55.1 & -11:44:38.4 & $16.04\pm0.09$ & $25.86\pm0.09$ & $0.79\pm0.04$ & $31.0\pm2.7$ & $0.82\pm0.01$\\
{\small \object{NGC4594-DGSAT-2}} & 12:39:51.6 & -11:20:24.8 & $19.28\pm0.08$ & $25.68\pm0.08$ & $0.57\pm0.02$ & $6.8\pm0.1$ & $0.85\pm0.01$\\
{\small \object{NGC4594-DGSAT-3}} & 12:39:32.8 & -11:13:38.5 & $17.77\pm0.11$ & $25.26\pm0.11$ & $0.19\pm0.07$ & $13.33\pm0.6$ & $0.82\pm0.04$\\
\hline
{\small \object{NGC4631-DGSAT-1}} & 12:42:53.1 & +32:27:19.0 & $18.74\pm0.08$ & $26.91\pm0.08$ & $0.73\pm0.03$ & $14.66\pm0.5$ & $0.81\pm0.01$\\
{\small \object{NGC4631-DGSAT-2}} & 12:42:06.1 & +32:37:14.8 & $17.25\pm0.08$ & $27.01\pm0.08$ & $0.48\pm0.02$ & $34.52\pm1.5$ & $0.74\pm0.01$\\
{\small \object{NGC4631-DGSAT-3}} & 12:41:08.0 & +32:26:50.4 & $20.95\pm0.09$ & $26.37\pm0.09$ & $0.28\pm0.03$ & $5.57\pm0.2$ & $0.63\pm0.02$\\
\hline
{\small \object{NGC7814-DGSAT-1}} & 0:03:24.0 & +16:11:13.6 &$18.05\pm0.06$ & $25.26\pm0.06$ & $0.38\pm0.05$ & $11.5\pm0.2$  & $0.71\pm0.01$\\
{\small \object{NGC7814-DGSAT-2}} & 0:03:06.8 & +16:18:33.3 &$18.19\pm0.05$ & $26.14\pm0.05$ & $0.66\pm0.02$ & $18.7\pm0.5$ & $0.43\pm0.01$\\

  \hline
  \end{tabular}
 \end{center}
\end{table*}

\subsection{The NGC 3628 field}\label{subsec:3628}
\begin{figure}
 \begin{center}
  \includegraphics[scale=0.35]{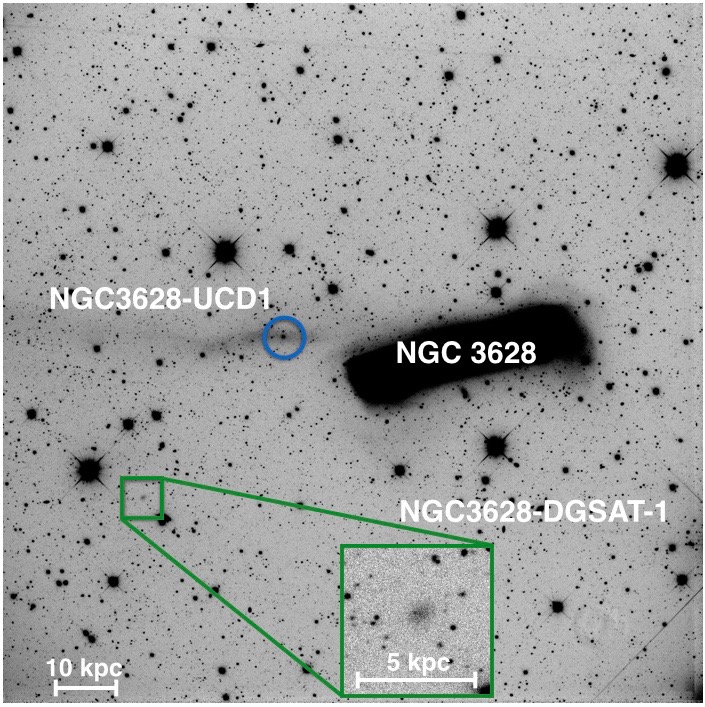}
  \caption{The $0^\circ\!\!.7 \times 0^\circ\!\!.7$ field of \object{NGC 3628}. North is up and east is to the left. The zoomed-in square shows the LSB galaxy we found in the image and the blue circle shows the \object{NGC3628-UCD1} \citep{jennings15}.\label{fig:ngc3628}}
  
 \end{center}
\end{figure}

\object{NGC 3628} is a spiral galaxy with an active nucleus and lies at an adopted distance of $10.9 \pm 2.4$ Mpc (Table \ref{t:galaxies}). Our image from the field of this galaxy comprises $0^\circ\!\!.7 \times 0^\circ\!\!.7$ and it is shown in Figure \ref{fig:ngc3628}. The pixel size of this image is 0.62 arcsec and its FWHM is $\approx2.7$ arcsec. The immediately observable features of this image are a heavily perturbed disk and a stream parallel to the disk of the galaxy. In the figure, we show the \object{NGC3628-UCD1}, which was identified and studied by \citet{jennings15} to be an embedded star cluster in the observed stream. The dwarf galaxy candidate that we found in this field is also shown in Figure \ref{fig:ngc3628}. The GALFIT modelling and its results are shown in Figure \ref{fig:dgsats} and Table \ref{t:results}.

\subsection{The NGC 4594 (\object{M104}) field}\label{subsec:4594}
\begin{figure}
 \begin{center}
  \includegraphics[scale=0.32]{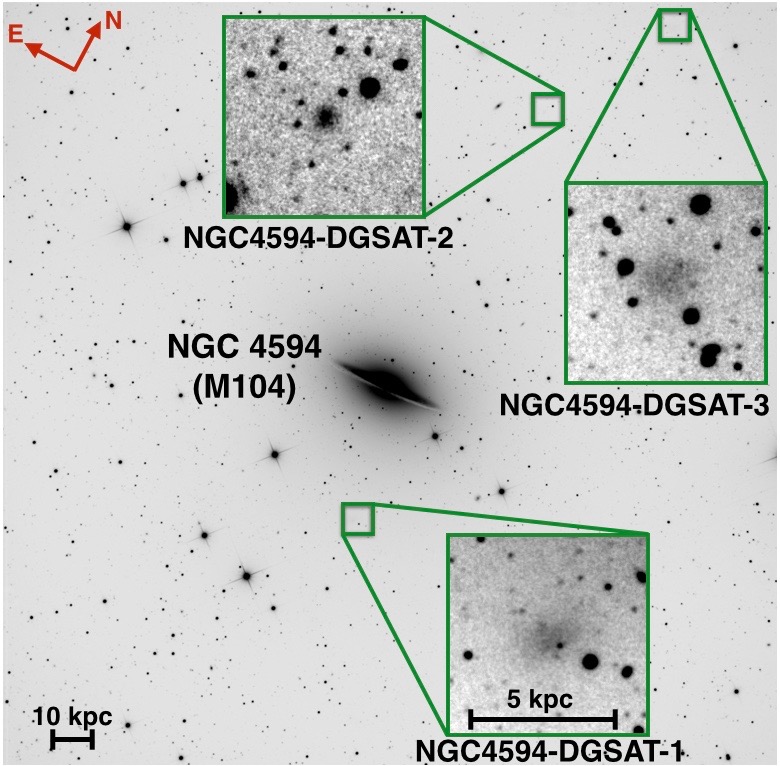}
  \caption{The $0^\circ\!\!.7 \times 0^\circ\!\!.7$ field of \object{NGC 4594}. North (N) and east (E) are indicated with arrows at the top left of the image. The zoomed-in squares show the LSB galaxies we found in the image.\label{fig:ngc4594}}
  
 \end{center}
\end{figure}

\object{NGC 4594} (also known as the Sombrero galaxy) is an unbarred galaxy with a large bulge and an active nucleus. The adopted distance to this galaxy is $11.1 \pm 4.1$ Mpc (see Table \ref{t:galaxies}). Our image of the field of this galaxy (Figure \ref{fig:ngc4594}) spans $0^\circ\!\!.7 \times 0^\circ\!\!.7$ and has a pixel size of 0.62 arcsec. We could detect three LSB galaxies in this image, which are also shown in Figure \ref{fig:ngc4594}. The results of GALFIT modelling for these three candidates are also shown in Figure \ref{fig:dgsats} and Table \ref{t:results}.\\

\subsection{The NGC 4631 field}\label{subsec:4631}
\begin{figure}
 \begin{center}
  \includegraphics[scale=0.19]{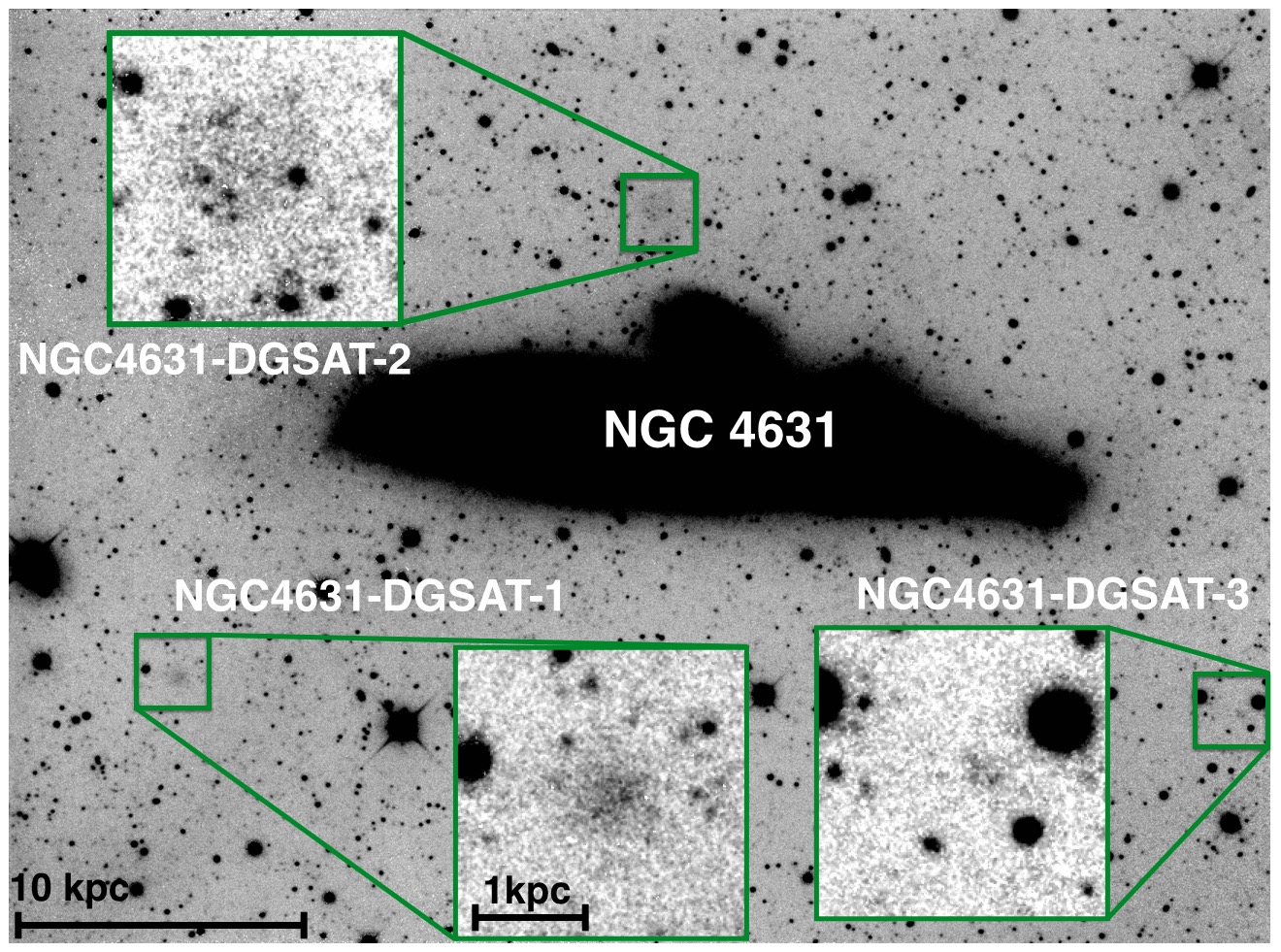}
  \caption{The $0^\circ\!\!.4 \times 0^\circ\!\!.3$ field of \object{NGC 4631}. North is up and east is to the left. The zoomed-in squares show the LSB galaxies in the image.\label{fig:ngc4631}}
  
 \end{center}
\end{figure}

\object{NGC 4631} (also known as the Whale Galaxy) is an edge-on galaxy at an adopted distance of $5.8 \pm 1.5$ Mpc (Table \ref{t:galaxies}). Our image of the field of this galaxy (Figure \ref{fig:ngc4631}) spans $0^\circ\!\!.4 \times 0^\circ\!\!.3$ and has a pixel size of 0.43 arcsec. \citet{karachentsev14} and \citet{delgado15} reported three dwarf galaxy candidates aligned with a stellar stream in deep images of this galaxy. Two of these LSB galaxies, which are in the field of our image, (and in our catalogue we refer to them as \object{NGC4631-DGSAT-1} and 2) are shown in Figure \ref{fig:ngc4631}. Our searching method could also find another new dwarf galaxy candidate in this field; \object{NGC4631-DGSAT-3}. The results of GALFIT modelling for these three candidates are shown in Figure \ref{fig:dgsats} and Table \ref{t:results}.\\

\subsection{The NGC 7814 field}\label{subsec:7814}
\begin{figure}
 \begin{center}
  \includegraphics[scale=0.45]{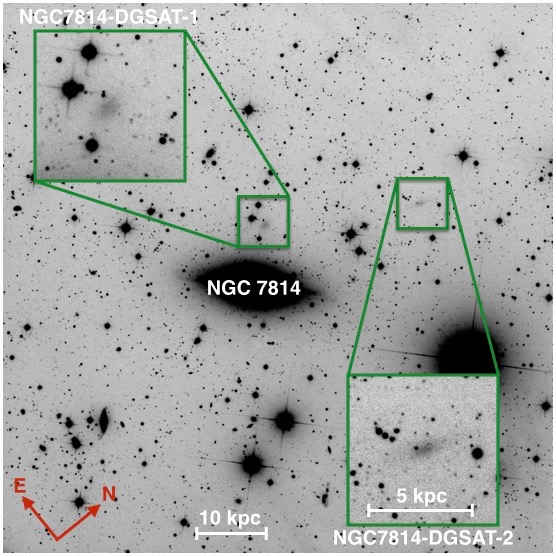}
  \caption{The $0^\circ\!\!.5 \times 0^\circ\!\!.5$ field of \object{NGC 7814}. North (N) and east (E) are indicated with arrows at the bottom left of the image. The zoomed-in squares show the LSB galaxies we found in the image.\label{fig:ngc7814}}
  
 \end{center}
\end{figure}

\object{NGC 7814} is an edge-on disk galaxy with a prominent bulge at the distance of $16.4 \pm 3.6$ Mpc (see Table \ref{t:galaxies}). Our image from the field of this galaxy spans $0^\circ\!\!.5 \times 0^\circ\!\!.5$ and is shown in Figure \ref{fig:ngc7814}. The pixel scale of this image is 0.43 arcsec/pixel and its FWHM is $\approx2.2$ arcsec. The two discovered LSB galaxies in the field of this galaxy, \object{NGC7814-DGSAT-1} and 2, are also shown in Figure \ref{fig:ngc7814}; they are visible even without zooming. The results of GALFIT modelling of these dwarf galaxy candidates are presented in Figure \ref{fig:dgsats}. The foreground stars were removed by fitting the PSF model of the image using DAOPHOT before using GALFIT. The clean residual images show that the models are good. The results can be seen in Table \ref{t:results}.

\section{Discussion}\label{sec:discussion}
All of the detected objects in this work have very low surface brightness (>25 mag/arcsec$^2$), and cannot be detected in the available images from large-scale surveys like the SDSS or PanSTARRs. This also makes it very difficult to undertake follow-up observations for obtaining their radial velocities (and confirming their association with the spiral galaxies) even for 8 meter class telescopes. For further analysis, and given their low angular projected distances, we assume that the discovered galaxies are the satellites of these nearby massive galaxies (see Table \ref{t:galaxies}), and derive their physical properties\footnote {In addition, the contamination of our search for faint satellites around massive spirals by background dE galaxies of similar colour is significantly reduced when we probe the surface brightness regime below 25 mag/arcsec$^2$, as is shown in the typical scaling relation diagram for early-type galaxies \citep[e.g. see the surface brightness versus size relation diagram in][]{toloba12}.}. Interestingly, some of our faint systems display insights of tidal disruption (e.g. NGC2683-DGSAT-1, NGC7814-DGSAT-2, and NGC5457-DF-4), which suggest their proximity to the spiral host.

\begin{figure}
 \begin{center}
  \includegraphics[scale=0.34]{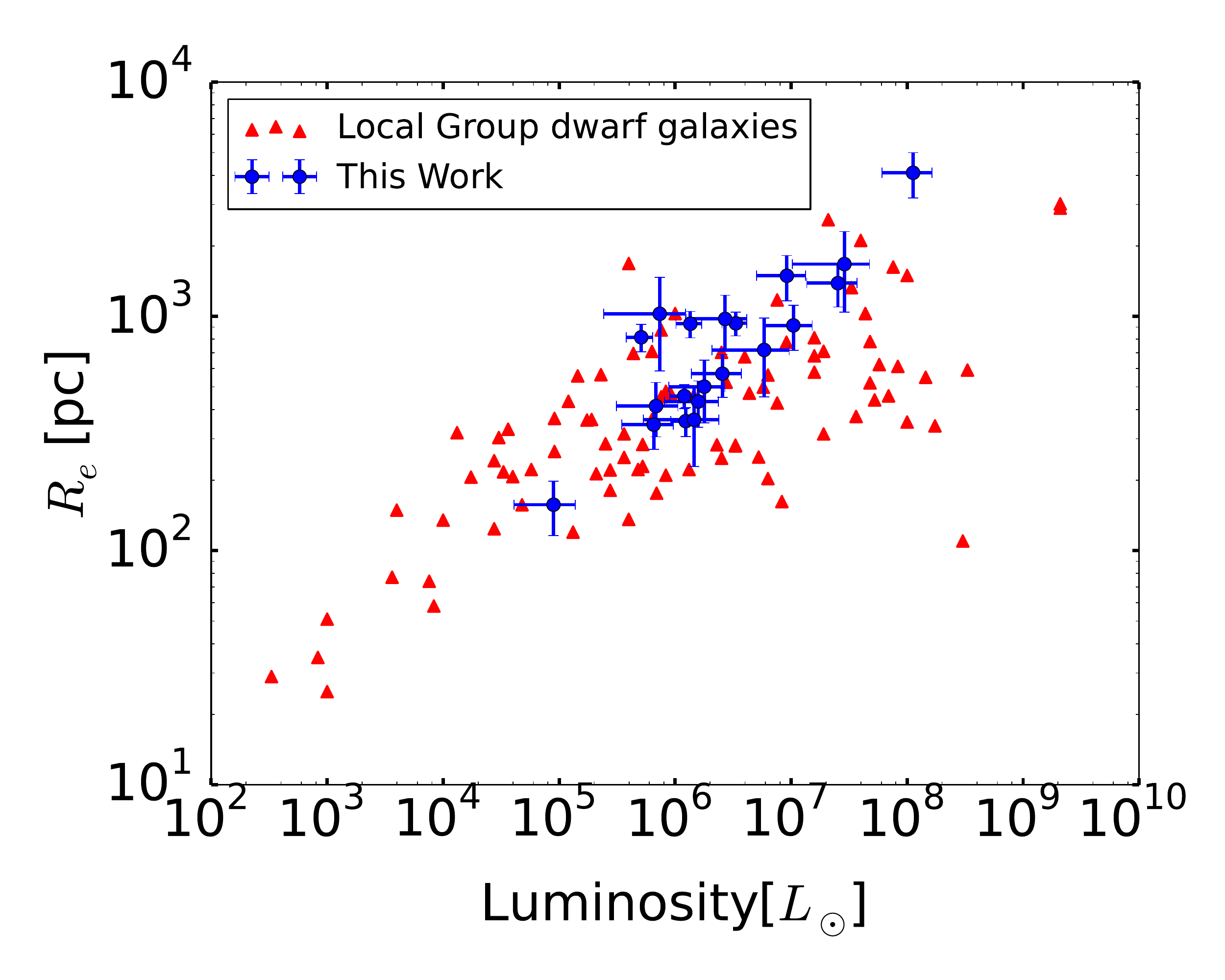}
  \includegraphics[scale=0.34]{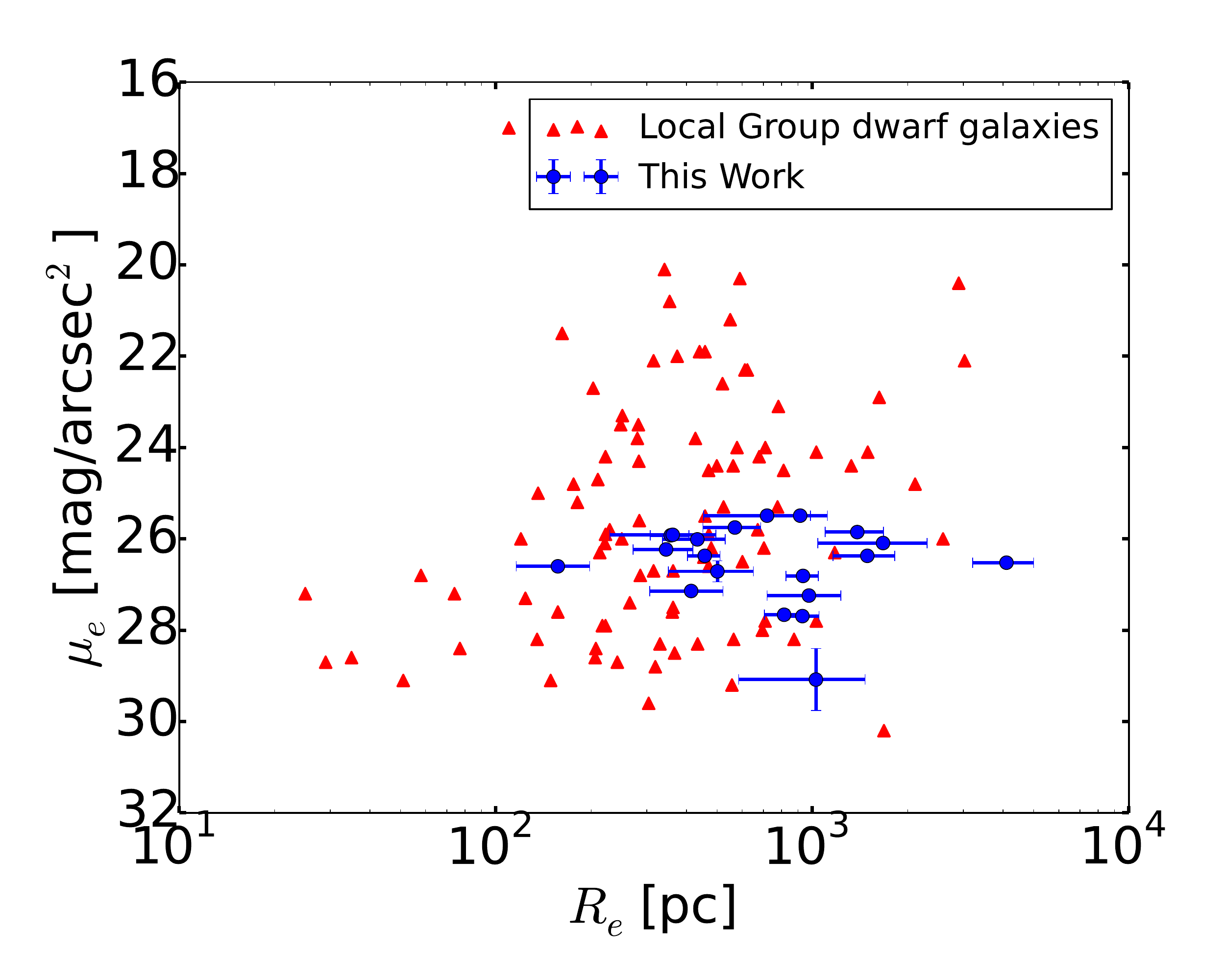}
  \includegraphics[scale=0.34]{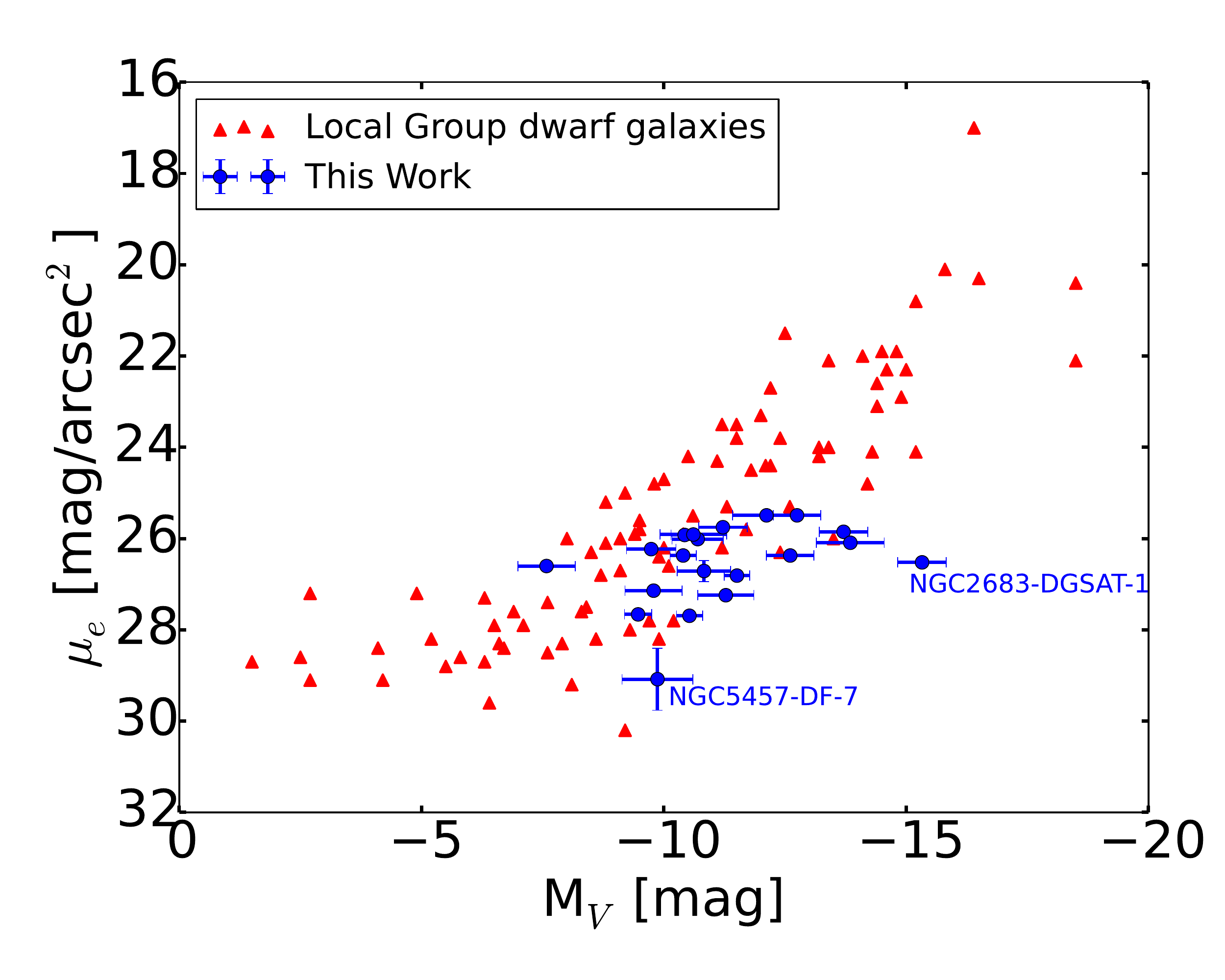}
  \caption{A comparison between the properties of the dwarf galaxy candidates studied in this work and the known dwarf galaxies of the Local Group \citep{mcconnachie12}. Top:  effective radius in pc vs. V band luminosity in  $L_{\odot}$. Middle: V band surface brightness in mag.arcsec$^{-2}$ vs. effective radius in pc. Bottom: V band surface brightness in mag.arcsec$^{-2}$ vs. absolute V band magnitude. In the latter, the outliers are emphasized by name. The V band quantities were obtained by transforming r band magnitudes to V magnitudes using \citet{fukugita96}. \label{fig:withLG}}
  
 \end{center}
\end{figure}

In Table \ref{t:results_phys} we list the effective radius (in pc), projected distance (in kpc) to their putative massive companion, absolute magnitude, M, (in mag) and luminosity, $L$, (in $10^6 \times L_{\odot}$) of all 20 dwarf galaxy candidates that we listed in Table \ref{t:results}.  Their r-band luminosities and their effective radii are in the ranges $0.1\times 10^6 \lesssim \left(\frac{L}{L_{\odot}}\right)_r\lesssim127 \times 10^6$, and 160 pc $\lesssim R_e \lesssim$ 4.1 kpc, respectively. Their r-band surface brightnesses and absolute magnitudes are in the ranges $25.3\lesssim\mu_{e}\lesssim28.8$ mag.arcsec$^{-2}$ and $-15.6 \lesssim M_r \lesssim -7.8$, respectively. Figure \ref{fig:withLG} shows the distribution of the dwarf galaxy candidates studied in this work in the $R_e$ vs. $L$, $\mu_e$ vs. $R_e$, and $\mu_e$ vs. M$_V$ planes. The known dwarf galaxies of the Local Group \citep{mcconnachie12} are also shown in these plots. In the $\mu_e$ vs. M$_V$ plane, two of our dwarf galaxy candidates, which are outliers to the overall trend, are indicated with names. NGC5457-DF-7 is very faint and its measured properties have large uncertainties, but NGC2683-DGSAT-1 is bright enough to show its irregular shape, which suggests that it might be undergoing tidal disruption.

As can be seen, the DGSAT candidates can be characterized by similar properties as those of the LG dwarf galaxies. This shows the ability of the DGSAT, and in general small telescopes, for discovering such faint systems in the Local Volume.
 
\begin{table*}
 \begin{center}
  \caption{Physical properties of the LSB galaxies assuming that they are satellites of the nearby (in projection) massive galaxies. $R_e$ and $d_P$ are effective radius and projected distance of the dwarf candidate to the centre of the host galaxy, respectively. $M_r$ denotes the absolute magnitude in r-band (assuming the distance moduli in Table \ref{t:galaxies}) and the last column provides the luminosity in units of $10^6$ solar luminosities, $L_{\odot}$.  \label{t:results_phys}}
  
  \begin{tabular}{lllll}
  
 ID & $R_e$ {\scriptsize(pc)} & $d_P$ {\scriptsize(kpc)}& M$_r$ {\scriptsize(mag)}&$\left(\frac{L}{L_{\odot}}\right)_r${\scriptsize $\times 10^6$}\\
  \hline
 {\small \object{NGC5457-DF-1}} & $457\pm54$ &$ 49.2 \pm 5.6 $ & $-10.63\pm0.27$ & $ 1.3 \pm 0.3 $ \\
 {\small \object{NGC5457-DF-3}} & $934\pm109$ &$ 87.4 \pm 10.0 $ & $-11.74\pm0.26$ & $ 3.8 \pm 0.9 $  \\
 {\small \object{NGC5457-DF-4}} & $930\pm120$ &$ 86.8 \pm 9.9 $ & $-10.76\pm0.27$ & $ 1.5 \pm 0.4 $ \\
 {\small \object{NGC5457-DF-6}} & $814\pm109$ &$ 133.3 \pm 15.3 $ & $-9.70\pm0.28$ & $ 0.6 \pm 0.1 $\\
 {\small \object{NGC5457-DF-7}} & $1026\pm441$ &$ 103.7 \pm 11.9 $ & $-10.10\pm0.73$ & $ 0.8 \pm 0.6 $\\
{\small \object{NGC5457-DGSAT-1}} & $357\pm50$ &$ 96.2 \pm 11.0 $ & $-10.66\pm0.28$ & $ 1.4 \pm 0.4 $ \\
\hline
{\small \object{NGC2683-DGSAT-1}} & $4103\pm898$ &$ 68.2 \pm 14.1 $ & $-15.56\pm0.50$ & $ 127.1 \pm 58.5 $\\
{\small \object{NGC2683-DGSAT-2}} & $1386\pm289$ &$ 104.9\pm 21.7 $ & $-13.94\pm0.50$ & $ 28.6 \pm 13.2 $\\
{\small \object{NGC2683-DGSAT-3}} & $569\pm118$ &$ 100.5 \pm 20.8 $ & $-11.45\pm0.50$ & $ 2.9 \pm 1.3 $\\
{\small \object{NGC2683-DGSAT-4}} & $501\pm150$ &$ 69.8 \pm 14.4 $ & $-11.06\pm0.55$ & $ 2.0 \pm 1.0 $\\
{\small \object{NGC2683-DGSAT-5}} & $345\pm74$ &$ 107.7 \pm 22.3 $ & $-9.97\pm0.51$ & $ 0.7 \pm 0.3 $\\
\hline
{\small \object{NGC3628-DGSAT-1}} & $433\pm97$ & $67.9 \pm 15.1$ & $-10.93\pm0.53$ & $ 1.8 \pm 0.9 $\\
\hline
{\small \object{NGC4594-DGSAT-1}} & $1672\pm631$ &$24.0 \pm 8.8$ & $-14.08\pm0.70$ & $ 32.5 \pm 21.0 $\\
{\small \object{NGC4594-DGSAT-2}} & $362\pm133$ &$54.9 \pm 20.2$ & $-10.84\pm0.69$ & $ 1.6 \pm 1.0 $\\
{\small \object{NGC4594-DGSAT-3}} & $719\pm266$ &$79.6 \pm 29.2$ & $-12.35\pm0.70$ & $ 6.6 \pm 4.3 $\\
\hline
{\small \object{NGC4631-DGSAT-1}} & $414\pm108$ &$19.3 \pm 5.0$ & $-10.02\pm0.59$ & $ 0.8 \pm 0.4 $\\
{\small \object{NGC4631-DGSAT-2}} & $975\pm256$ &$8.1 \pm 2.1$ & $-11.51\pm0.58$ & $ 3.0 \pm 1.6 $\\
{\small \object{NGC4631-DGSAT-3}} & $157\pm41$ &$22.5 \pm 5.8$ & $-7.81\pm0.59$ & $ 0.1 \pm 0.05 $\\
\hline
{\small \object{NGC7814-DGSAT-1}} & $915\pm200$ &$16.2 \pm 3.5$ & $-12.98\pm0.49$ & $ 11.8 \pm 5.3 $\\
{\small \object{NGC7814-DGSAT-2}} & $1491\pm328$ &$47.6 \pm 10.4$ & $-12.84\pm0.49$ & $ 10.4 \pm 4.7 $\\

  \hline

  \end{tabular}
 \end{center}
\end{table*}

\section{Conclusion}\label{sec:conclusion}
We presented the first results of the DGSAT project and its ability to find LSB galaxies around nearby Milky Way-like galaxies using a network of robotic amateur telescopes. We developed a semi-automatic pipeline to calibrate the luminance images taken by amateur telescopes, search for dwarf galaxy candidates, and extract their observed parameters. By exploring the fields of six nearby massive galaxies NGC 2683, NGC 3628, NGC 4594 (M104), NGC 4631, NGC 5457 (M101), and NGC 7814, we discovered eleven so far unknown LSB galaxies in our images. The models of these galaxies were obtained using the GALFIT software and by fitting a Sersic function to their light profile. While revisiting the field of M101, we have discovered a new LSB galaxy that was not reported by the Dragonfly team, which recently observed the field of this galaxy. Our results for the rest of the LSB galaxies in the field of M101 confirm those of the Dragonfly study but our image provides significantly better angular resolution.

The LSB galaxies in the fields of the other mentioned massive galaxies show similar observed properties to those in the field of M101. In addition to the eleven newly identified dwarf galaxy candidates, we also analysed the morphology of nine already known objects. The surface brightness of all of these galaxies are in the range $25.3\lesssim\mu_{e}\lesssim28.8$ mag.arcsec$^{-2}$ and their Sersic indices are $n\lesssim 1,$ which are similar to those of the dwarf satellites in the Local Group \citep{mcconnachie12}. Assuming that they are dwarf satellites of their neighbouring (in projection) massive galaxies, their r-band absolute magnitude, their luminosities, and their effective radii are in the ranges $-15.6 \lesssim M_r \lesssim -7.8$, $0.1\times 10^6 \lesssim \left(\frac{L}{L_{\odot}}\right)_r\lesssim127 \times 10^6$, and 160 pc $\lesssim R_e \lesssim$ 4.1 kpc, respectively. In particular, DGSAT is able to detect LSB systems with similar observed properties to those of the ``classical'' dwarf spheroidal galaxies around the Milky Way.

To confirm that the discovered galaxies are dwarf satellites of their nearby massive galaxies, further observations are required. Our results show the potential of amateur telescopes in discovering more dwarf galaxies in the Local Volume, which enables us to further test models of galaxy formation and evolution outside the Local Group.

\begin{acknowledgements}
We thank the referee for providing constructive comments on the manuscript. BJ thanks Chien Peng for his kind help with the GALFIT software, Eva Grebel for her constructive comments on the project, and Luca Fossati and Tim Schrabback for their useful guidance on image analysis. BJ was supported for this research through stipends from the International Max Planck Research School (IMPRS) for Astronomy and Astrophysics at the Universities of Bonn and Cologne, and from Karl Menten and SPODYR groups. DMD acknowledges support by the Sonderforschungsbereich (SFB) 881 "The Milky Way system" of the German Research Foundation (DFG), particularly through the sub-project A2. This research has made use of the SIMBAD database, and the VizieR catalogue access tool, operated at CDS in Strasbourg, France, and the "Aladin sky atlas", which was developed at the same location.  
\end{acknowledgements}

\bibliography{biblio}

\end{document}